\documentclass[prl,twocolumn,amssymb,showpacs,amsmath,nobibnotes,
superscriptaddress,floatfix,aps]{revtex4}
\usepackage{graphicx}
\usepackage{bm}

\renewcommand{\case}[2]{\ensuremath{{\textstyle\frac{#1}{#2}}}}
\newcommand{\aXoarse}{0.175}
\newcommand{\aCoarse}{0.121}
\newcommand{\aFine}{0.086}

\begin{document}
\title{Charmed-Meson Decay Constants in Three-Flavor Lattice QCD}
\author{C.~Aubin}
\affiliation{Physics Department, Columbia University,
New York, New York, USA}
\author{C.~Bernard} 
\affiliation{Department of Physics, Washington University,
St.~Louis, Missouri, USA}
\author{C.~DeTar} 
\affiliation{Physics Department, University of Utah,
Salt Lake City, Utah, USA}
\author{M.~Di Pierro}
\affiliation{School of Computer Science, Telecommunications and Information
Systems, DePaul University, Chicago, Illinois, USA}
\author{E. D. Freeland}
\affiliation{Liberal Arts Department, The School of the Art Institute
of Chicago, Chicago, Illinois, USA}
\author{Steven Gottlieb} 
\affiliation{Department of Physics, Indiana University,
Bloomington, Indiana, USA}
\author{U.~M.~Heller}
\affiliation{American Physical Society, % One Research Road, Box 9000, 
Ridge, New York, USA}
\author{J.~E.~Hetrick}
\affiliation{Physics Department, University of the Pacific,
Stockton, California, USA}
\author{A.~X.~El-Khadra}
\affiliation{Physics Department, University of Illinois,
Urbana, Illinois, USA}
\author{A.~S.~Kronfeld}
\affiliation{Fermi National Accelerator Laboratory, Batavia, Illinois, USA}
\author{L.~Levkova}
\affiliation{Department of Physics, Indiana University,
Bloomington, Indiana, USA}
\author{P.~B.~Mackenzie} 
\affiliation{Fermi National Accelerator Laboratory, Batavia, Illinois, USA}
\author{D.~Menscher}
\affiliation{Physics Department, University of Illinois,
Urbana, Illinois, USA}
\author{F. Maresca}
\affiliation{Physics Department, University of Utah,
Salt Lake City, Utah, USA}
\author{M.~Nobes}
\affiliation{Laboratory of Elementary-Particle Physics, Cornell University,
Ithaca, New York, USA}
\author{M.~Okamoto}
\affiliation{Fermi National Accelerator Laboratory, Batavia, Illinois, USA}
\author{D.~Renner}
\affiliation{Department of Physics, University of Arizona,
Tucson, Arizona, USA}
\author{J.~Simone}
\affiliation{Fermi National Accelerator Laboratory, Batavia, Illinois, USA}
\author{R.~Sugar}
\affiliation{Department of Physics, University of California,
Santa Barbara, California, USA}
\author{D.~Toussaint}
\affiliation{Department of Physics, University of Arizona,
Tucson, Arizona, USA}
\author{H.~D.~Trottier}
\affiliation{Physics Department, Simon Fraser University,
Burnaby, British Columbia, Canada}
\collaboration{Fermilab Lattice, MILC, and HPQCD Collaborations}
\noaffiliation

\date{September 8, 2005}
\pacs{13.20.Fc,12.38.Gc}
\begin{abstract}
We present the first lattice QCD calculation with realistic sea quark
content of the $D^+$-meson decay constant $f_{D^+}$.
We use the MILC Collaboration's publicly available ensembles of lattice
gauge fields, which have a quark sea with two flavors (up and down)
much lighter than a third (strange).
We obtain $f_{D^+} = 201 \pm 3 \pm 17~\textrm{MeV}$, where the errors
are statistical and a combination of systematic errors.
We also obtain $f_{D_s} = 249 \pm 3 \pm 16~\textrm{MeV}$ for the
$D_s$~meson.
\end{abstract}

\maketitle

Flavor physics currently plays a central role in elementary particle
physics~\cite{ckm:2005}.
To aid the experimental search for physics beyond the standard model,
several hadronic matrix elements must be calculated nonperturbatively
from quantum chromodynamics (QCD).
One of the most important of these is the decay constant of the 
$B$~meson $f_B$~\cite{Lubicz:2004nn}.
Any framework for calculating~$f_B$ should, therefore, be subjected to 
stringent tests, and such a test is a key aim of this Letter.
\vskip -1pt

The most promising method for these nonperturbative calculations is
numerical lattice~QCD.
For many years the results suffered from an unrealistic treatment of the
effects of sea quarks.
In the last few years, however, this obstacle seems to have been
removed: with three flavors of sea quarks lattice QCD now agrees with
experiment for a wide variety of hadronic
quantities~\cite{Davies:2003ik}.
This validation of lattice QCD has been realized, so far, only for 
so-called ``gold-plated'' quantities: 
masses and matrix elements of the simplest hadronic states.
Note, however, that many of the hadronic matrix elements relevant to 
flavor physics are in this class, including~$f_B$.
\vskip -1pt

The  challenges in computing~$f_B$ are essentially the same for
the $D^+$-meson decay constant~$f_{D^+}$.
Experiments have observed the leptonic decay 
${D^+}\to l^+\nu_l$, but not $B^+\to l^+\nu_l$.
One can, thus, determine $|V_{cd}|f_{D^+}$, where $V_{cd}$ is an element
of the Cabibbo-Kobayashi-Maskawa (CKM) matrix.
Taking $|V_{cd}|$ from elsewhere, one gets~$f_{D^+}$.
In 2004 the CLEO-c Collaboration measured $f_{D^+}$ with a 20\%
error~\cite{Bonvicini:2004gv}, and a more precise measurement is
expected soon.

This Letter reports the first lattice-QCD calculation of $f_{D^+}$ with
three flavors of sea quarks~\cite{Simone:2004fr}.
We find
\begin{equation}
	f_{D^+} = 201 \pm  3 \pm  6 \pm  9 \pm 13~\textrm{MeV},
	\label{eq:fD+}
\end{equation}
where the uncertainties are statistical,
and a sequence of systematic effects, discussed below.
We also obtain the decay constant of the $D_s$ meson,
\begin{equation}
	f_{D_s} = 249 \pm  3 \pm  7 \pm 11 \pm 10~\textrm{MeV}.
	\label{eq:fDs} 
\end{equation}
The second result is more precise than a recent lattice-QCD calculation
with the same sea quark content but non-relativistic heavy quarks, which
found
$f_{D_s}=290\pm 20\pm 41~\textrm{MeV}$~\cite{Wingate:2003gm}.
These results are more reliable than older
calculations~\cite{El-Khadra:1997hq} because we now incorporate (three)
sea quarks and, for $f_{D^+}$, also because the light valence quark
masses are smaller than before.

These results test the methods of Ref.~\cite{Davies:2003ik} 
because they are predictions.
The input parameters have been fixed 
previously~\cite{Davies:2003ik,Aubin:2004ck,Aubin:2004fs,%
diPierro:2003iw,diPierro:2003bu},
and, once comparably precise experimental measurements become
available, one can see how Eqs.~(\ref{eq:fD+}) and~(\ref{eq:fDs}) fare.
Indeed, this work is part of a program to calculate matrix elements for
leptonic and semileptonic
decays~\cite{diPierro:2003iw,Aubin:2004ej,Okamoto:2004xg}, neutral-meson
mixing, and quarkonium~\cite{diPierro:2003bu,Allison:2004be}.
So far, these lattice QCD calculations agree with experiment for the
normalization of $D$-meson semileptonic form
factors~\cite{Aubin:2004ej,Ablikim:2004ej,Huang:2004fr}.
They also have predicted correctly the form-factor
shape~\cite{Aubin:2004ej,Link:2004dh},
as well as the mass of the $B_c$
meson~\cite{Allison:2004be,Acosta:2005us}.
\vskip -1pt

In this set of calculations we use ensembles of unquenched lattice gauge
fields generated by the MILC
Collaboration~\cite{Bernard:2001av,Aubin:2004fs},
with lattice spacing $a=\aXoarse$, \aCoarse, and \aFine~fm.
The key feature of these ensembles is that they incorporate three
flavors of sea quarks, one whose mass is close to that of the strange
quark, and two with a common mass taken as light as possible.
\vskip -1pt

For the sea quark and light valence quark we use the ``Asqtad''
staggered-fermion action~\cite{asqtad}.
Several different quark masses appear in this calculation;
for convenience, they are defined in Table~\ref{tbl:masses}.
\begin{table}
	\centering
	\caption[tbl:masses]{Notation for quark masses used in this Letter.}
	\label{tbl:masses}
	\begin{tabular*}{0.48\textwidth}{c@{\extracolsep{\fill}}l@{\extracolsep{\fill}}l}
		\hline\hline
		$m$ & \multicolumn{1}{c}{Description} & \multicolumn{1}{c}{Remark} \\
		\hline
		$m_c$ & Charmed                      quark & 
			From $m_{D_s}$~\cite{diPierro:2003iw,diPierro:2003bu}  \\
		$m_s$ & Physical     strange         quark & 
			From $m_K^2$~\cite{Aubin:2004ck} \\
		$m_u$ & Physical        up           quark & 
			$m_u = m_s/45.5$~\cite{Aubin:2004fs}  \\
		$m_d$ & Physical       down          quark & 
			$m_d = m_s/19.6$~\cite{Aubin:2004fs}  \\
		$m_h$ & Simulation's heavier   sea   quark & 
			$m_h\approx1.1m_s$  \\
		$m_l$ & Simulation's lighter   sea   quark & 
			$0.1m_s\leq m_l\lesssim 0.8m_s$ \\
		$m_q$ & Simulation's  light  valence quark & 
			$0.1m_s\leq m_q\lesssim m_s$ \\
		\hline\hline
	\end{tabular*}
\end{table}
At $a=\aXoarse$, \aCoarse, and \aFine~fm there are, respectively, 4,~5, and~2
ensembles with various sea quark masses
$(m_l,m_h)$~\cite{Bernard:2001av,Aubin:2004fs}.
The larger simulation mass, $m_h$ is close to the physical 
strange quark mass~$m_s$.
The light pair's mass $m_l$ is not as small as those of the up and
down quark in Nature, but the range $0.1m_s\leq m_l\lesssim 0.8m_s$
suffices to control the extrapolation in quark mass with chiral
perturbation theory ($\chi$PT).
For carrying out the chiral extrapolation, it is useful to
allow the valence mass~$m_q$ to vary separately from the sea
mass~\cite{Bernard:1993sv}.
At $a=\aXoarse$, \aCoarse, and \aFine~fm we have, respectively,
6, 12, and 8 or 5 values of the valence mass,
in the range $0.1m_s\leq m_q\lesssim m_s$.
\vskip -1pt

A drawback of staggered fermions is that they come in four species, 
called tastes.
The steps taken to eliminate three extra tastes per flavor are not (yet)
proven, although there are several signs that they are valid.
Calculations of $f_{D^+}$ and $f_{D_s}$ are sensitive to these steps: 
if Eqs.~(\ref{eq:fD+}) and~(\ref{eq:fDs}) agree with precise measurements,
it should be more plausible
that the techniques used to reduce four tastes to one are correct.

For the charmed quark we use the Fermilab action for heavy
quarks~\cite{El-Khadra:1996mp}.
Discretization effects are entangled with the heavy-quark expansion,
so we use heavy-quark effective theory (HQET) as a theory of cutoff
effects~\cite{Kronfeld:2000ck}.
This provides good control, as discussed in Ref.~\cite{Kronfeld:2003sd},
and the framework has been tested with the (successful) prediction of
the $B_c$ meson mass~\cite{Allison:2004be}.
Nevertheless, heavy-quark discretization effects are the largest source
of systematic error in~$f_{D_s}$, and the second-largest in~$f_{D^+}$.

The decay constant $f_{D_q}$, for a $D_q$ meson with light valence
quark~$q$ and momentum $p_\mu$, is defined by~\cite{Eidelman:wy}
\begin{equation}
	\langle0|A_\mu|D_q\rangle = if_{D_q}p_\mu ,
% 	= \phi_{D_q} v_\mu \sqrt{m_{D_q}} ,
	\label{eq:0AD}
\end{equation}
% $\langle0|A_\mu|D_q\rangle = if_{D_q}p_\mu$,
where $A_\mu=\bar{q}\gamma_\mu\gamma_5c$ is an electroweak axial vector
current. % and $v_\mu=p_\mu/m_{D_q}$.
The combination $\phi_q=f_{D_q}\sqrt{m_{D_q}}$ emerges directly from the 
lattice Monte Carlo calculations.
As usual in lattice gauge theory, we compute two-point correlation
functions
$C_2(t)=\langle O_{D_q}^\dagger(t) O_{D_q}(0) \rangle$,
$C_A(t)=\langle  A_4           (t) O_{D_q}(0) \rangle$,
where $O_{D_q}$ is an operator with the quantum numbers of the charmed 
pseudoscalar meson, and $A_4$ is the (lattice) axial vector current.
The operators are built from the heavy-quark and staggered-quark fields
as in Ref.~\cite{Wingate:2002fh}.
We extract the $D_q$ mass and the amplitudes
$\langle D|O_{D_q}|0\rangle$ and $\langle 0|A_4|D\rangle$
from fits to the known $t$ dependence.
Statistical errors are determined with the bootstrap method, which
allows us to keep track of correlations.

The lattice axial vector current must be multiplied by a renormalization
factor~$Z_{A^{cq}_4}$.
We write~\cite{El-Khadra:2001rv}
$Z_{A^{cq}_4}=\rho_{A^{cq}_4}(Z_{V^{cc}_4} Z_{V^{qq}_4})^{1/2}$,
because the flavor-conserving renormalization factors $Z_{V^{cc}_4}$ and
$Z_{V^{qq}_4}$ are easy to compute nonperturbatively.
The remaining factor $\rho_{A^{cq}_4}$ should be close to unity
because the radiative corrections mostly cancel~\cite{Harada:2001fi}.
A one-loop calculation gives~\cite{one-loop} 
% latticeÊ rho_a4 error=alpha_s(rho-1)
% ÊÊ FÊÊÊÊ 1.0319 0.0072
% ÊÊ CÊÊÊÊ 1.0438 0.0102
% ÊÊ XÊÊÊÊ 1.0524 0.0137
$\rho_{A^{cq}_4}=1.052$, 1.044, and 1.032
at $a=\aXoarse$, \aCoarse, and \aFine~fm.
We estimate the uncertainty of higher-order corrections to be
$2\alpha_s(\rho_{A^{cq}_4}-1)\approx1.3\%$; 
$\alpha_s$ is the strong coupling.

The heart of our analysis is the chiral extrapolation, from the 
simulated to the physical quark masses.
It is necessary, and non-trivial, because the cloud of ``pions''
surrounding the simulated $D_q$ mesons is not the same as for real
pions.
With staggered quarks the (squared) pseudoscalar meson masses are
\begin{equation}
	M^2_{ab,\xi} = (m_a+m_b)\mu + a^2\Delta_\xi,
	\label{eq:mesonmass}
\end{equation}
where $m_a$ and $m_b$ are quark masses, $\mu$ is a parameter of $\chi$PT, 
and the representation of the meson under the taste symmetry group is
labeled by $\xi=P, A, T, V, I$ \cite{Lee:1999zx}.
A symmetry as $m_a,m_b\to0$ ensures that $\Delta_P=0$.
The ``pion'' cloud in the simulation includes all these pseudoscalars.

According to next-to-leading order $\chi$PT the decay constant takes 
the form
\begin{equation}
	\phi_q = \Phi\left[1 + \Delta f_q(m_q,m_l,m_h) + 
		p_q(m_q,m_l,m_h) \right],
	\label{eq:phiq}
\end{equation}
where $\Phi$ is a quark-mass-independent parameter.
$\Delta f_q$ arises from loop processes involving light pseudoscalar 
mesons, and $p_q$ is an analytic function.
To obtain them one must take into account the
flavor-taste symmetry of the simulation~\cite{Lee:1999zx} and the
inequality (in general) of the valence and sea quark
masses~\cite{Bernard:1993sv}.
One finds~\cite{Aubin:2004xd}
\begin{equation}
	\Delta f_q = - \frac{1+3g^2}{2(4\pi f_\pi)^2}\left[
		\bar{h}_q + h_q^I + 
		a^2\left(\delta'_Ah_q^A + \delta'_Vh_q^V \right)\right],
\end{equation}
where $f_\pi\approx131$~MeV is the pion decay constant,
$g$ is the $D$-$D^*$-$\pi$ coupling~\cite{Grinstein:1992qt},
and $\delta'_A$, $\delta'_V$ parametrize effects that arise 
only at non-zero lattice spacing~\cite{Lee:1999zx}.
The terms $\bar{h}_q$, $h^I_q$, $h^A_q$, and $h^V_q$ are functions of
the pseudoscalar meson masses.
The last two, $h^A_q$ and $h^V_q$, are too cumbersome to write out here.
It is instructive to show the other two, $\bar{h}_q$ and $h^I_q$,
when $m_q=m_l$ or $m_h$: % and assuming $m_h=m_s$,
\begin{eqnarray}
	\bar{h}_q & = & \case{1}{16}{\textstyle\sum_\xi} n_\xi \left[
		2 I(M^2_{ql,\xi}) + I(M^2_{qh,\xi}) \right],
		\label{eq:hq}  \\
	h_l^I & = & - \case{1}{2} I(M^2_{ll,I}) + 	
	\case{1}{6}I(M^2_{\eta,I}),
		\label{eq:hIq}  \\
	h_h^I & = & - I(M^2_{hh,I}) + \case{2}{3}I(M^2_{\eta,I}),
		\label{eq:hIs}
\end{eqnarray}
where $I(M^2)=M^2\ln M^2/\Lambda^2_\chi$
(with $\Lambda_\chi$ the chiral scale), and 
$M^2_{\eta,I}=(M^2_{ll,I}+2M^2_{hh,I})/3$.
The term $h^I_q$ receives contributions only from taste-singlet mesons 
(representation~$I$).
The term $\bar{h}_q$ receives contributions from all representations,
with multiplicity $n_\xi=1,4,6,4,1$ for $\xi=P,A,T,V,I$, respectively.
The analytic function is
\begin{equation}
	p_q = (2m_l+m_h)f_1(\Lambda_\chi) + m_q f_2(\Lambda_\chi) + O(a^2),
	\label{eq:pq}
\end{equation}
where $f_1$ and $f_2$ are quark-mass-independent parameters.
They are essentially couplings of the chiral Lagrangian, and their 
$\Lambda_\chi$ dependence must cancel that of $\Delta f_q$.
This specifies $O(a^2)$ terms proportional to $f_1$ and $f_2$, which can
be removed after our fit.
We estimate the remaining $O(a^2)$ effects of light quarks to be small: 
around~4\% at $a=\aCoarse$~fm and $1.4\%$ at $a=\aFine$~fm.

The salient feature~\cite{Kronfeld:2002ab} of the chiral extrapolation
of $\phi_q$ is that $\Delta f_q$ contains a ``chiral log''
$I(2m_q\mu)\sim m_q\ln m_q$, which has a characteristic curvature as
$m_q\to 0$.
Equations~(\ref{eq:mesonmass})--(\ref{eq:hIq}) show that 
the chiral log is diluted by discretization effects,
because $a^2\Delta_\xi\neq0$ for $\xi\neq P$. % $\xi=A,T,V,I$.

We can now discuss how we carry out the chiral extrapolation.
Recall that we compute $\phi_q$ for many combinations of the valence and
light sea quark masses.
At each lattice spacing, we fit all results for $\phi_q$ to the mass
dependence prescribed by Eqs.~(\ref{eq:mesonmass})--(\ref{eq:pq}).
Of the twelve parameters, eight---$\mu$, the four non-zero
$\Delta_\xi$, $f_\pi$, $\delta'_A$, and $\delta'_V$---appear in the
$\chi$PT for light pseudoscalar mesons.
We constrain them with prior distributions whose central value
and width are taken from the $\chi$PT analysis of pseudoscalar meson
masses and decay constants on the same ensembles of lattice gauge
fields~\cite{Aubin:2004fs}.
The rest---$\Phi$, $g^2$, $f_1$, and $f_2$---appear only for 
charmed mesons.
We constrain $g^2$ to its experimentally measured value, within its
measured uncertainty~\cite{Anastassov:2001cw}.
Thus, only three parameters---$\Phi$, $f_1$, and $f_2$---are determined
solely by the $\phi_q$~fit.
To obtain physical results we reconstitute the fit setting 
the light sea quark mass $m_l\to(m_u+m_d)/2$, and
$\Delta_\xi=\delta'_{A,V}=0$.
For $\phi_d$ ($\phi_s$) we set the light valence mass 
$m_q\to m_d$~($m_s$). 

\begin{figure}
	\includegraphics[width=0.44\textwidth]{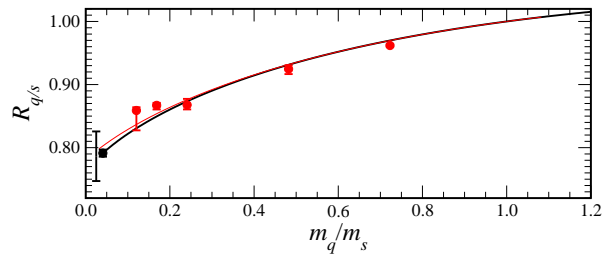}
	\caption[fig:chiral:D+]{Chiral extrapolation of $R_{q/s}$ at 
	$a=\aCoarse$~fm.  Data points show only statistical errors, 
	but the systematic error of fitting is shown at left.}
	\label{fig:chiral:D+}
\end{figure}

To isolate the uncertainties of the chiral extrapolation from other 
sources of uncertainty, we consider the ratio
$R_{q/s}=\phi_q/\phi_s$.
Figure~\ref{fig:chiral:D+} shows $R_{q/s}$ at $a=\aCoarse$~fm
as a function of $m_q/m_s$, projected onto $m_q=m_l$.
The gray (red) curve is the result of the full fit of $\phi_q$
to the separate sea- and valence-mass dependence.
The black curve, and the extrapolated value at $m_q/m_s=0.05$, results
from setting $\Delta_\xi=\delta'_{A,V}=0$ when reconstituting the fit.
At the other lattice spacings we obtain similar results.

The precision after the chiral extrapolation is, however, a bit illusory.
We tried several variations in the fit procedure:
fitting the ratio directly;
adding terms quadratic in the quark masses to Eq.~(\ref{eq:pq});
variations in the widths of the prior constraints of the parameters.
When these possibilities are taken into account, the extrapolated value 
of $R_{d/s}$ varies by 5\%, which we take as a systematic uncertainty.
This variation could be reduced with higher statistics at the lightest
sea quark masses.

\begin{figure}
	\centering
	\includegraphics[width=0.44\textwidth]{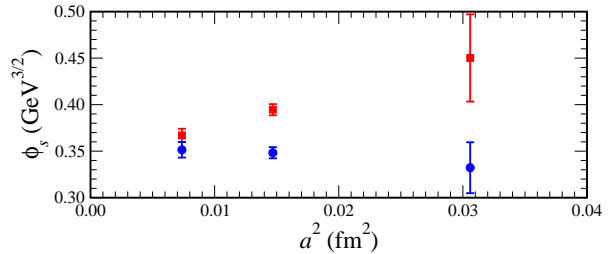}
	\caption[fig:lat:Ds]{Dependence of $\phi_s$ on~$a^2$.
	Circles result from removing the $O(a^2)$ pieces in
	Eq.~(\ref{eq:pq}); squares omit this step.}
	\label{fig:lat:Ds}
\end{figure}

The lattice spacing dependence of $\phi_s=f_{D_s}\sqrt{m_{D_s}}$ is
shown in Fig.~\ref{fig:lat:Ds}.
The (blue) circles are the main results.
In a preliminary report of this work~\cite{Simone:2004fr}
the $O(a^2)$ terms in $\phi_s$ were not removed.
The (red) squares illustrate the effect of omitting this step.
As one can see, the effect is small at $a=\aFine$~fm, but it is the
main reason why the results in Eqs.~(\ref{eq:fD+})
and~(\ref{eq:fDs}) are smaller than in Ref.~\cite{Simone:2004fr}.

The $\chi$PT expressions for $\phi_q$ assume that the $D_q$ meson is 
static.
Since its mass is around 1900~MeV and the pseudoscalars are a few 
hundred MeV, this is a good starting point.
Some corrections to this approximation can be absorbed into the fit 
parameters, with no real change in the analysis.
A more interesting change arises in the one-loop self-energy diagrams, 
for which the function $I(M^2)$ is modified, and depends 
on $m_{D^*}-m_D$ as well as $M$.
By replacing our standard  extrapolation by one using the modified function,
we estimate the associated error to be 1.5\% or less.
Finite-volume effects also modify $I(M^2)$: based on our experience with
$f_\pi$ and $f_K$~\cite{Aubin:2004fs} and on continuum $\chi$PT
\cite{Arndt:2004bg}, we estimate a further error of 1.5\% or~less.

Although $\chi$PT is able to remove (most of) the light-quark 
discretization errors, heavy-quark discretization effects remain.
We estimate this uncertainty using HQET as a theory of cutoff 
effects~\cite{Kronfeld:2000ck,Kronfeld:2003sd}.
To arrive at a numerical estimate, one must choose a typical 
scale~$\bar{\Lambda}$ for the soft interactions;
we choose $\bar{\Lambda}\approx500$--$700$~MeV.
We then estimate a discretization uncertainty of 2.7--4.2\% at 
$a=\aFine$~fm.
Similarly, the results at $a=\aCoarse$~fm are expected to lie within 1--2\% 
of those at $a=\aFine$~fm.

Because we cannot disentangle heavy- and light-quark discretization 
effects, to quote final results we average the results at $a=\aFine$ and 
\aCoarse~fm.
We then find 
\begin{eqnarray}
% 	R_{d/s} & = & 0.7864 \pm 0.0040 \pm 0.0047 \pm 0.0039 \pm 0.0423
	R_{d/s} & = & 0.786(04)(05)(04)(42)
		\label{eq:R} \\
%   \phi_s  & = & 0.3493 \pm 0.0049 \pm 0.0098 \pm 0.0154 \pm 0.0138
	\phi_s  & = & 0.349(05)(10)(15)(14)~\textrm{GeV}^{3/2} , 
		\label{eq:phis}
\end{eqnarray}
which are the principal results of this work.
The uncertainties (in parentheses) are, respectively, from
statistics,
input parameters $a$ and $m_c$,
heavy-quark discretization effects, and
chiral extrapolation.
A full error budget is in Table~\ref{tbl:errors};
\begin{table}
	\centering
	\caption[tbl:errors]{Error budget (in per cent)
		for $R_{d/s}$, $\phi_s$, $\phi_d$.}
	\label{tbl:errors}
	\begin{tabular*}{0.48\textwidth}{l@{\extracolsep{\fill}}r@{\extracolsep{\fill}}r@{\extracolsep{\fill}}r}
		\hline\hline
		source & $R_{d/s}$ & $\phi_s$ & $\phi_d$  \\
		\hline
		statistics & 0.5 & 1.4 & 1.5 \\
		input parameters $a$ and $m_c$ & 0.6 & 2.8 & 2.9 \\
%         \quad tuning $m_c$ & 0.2 & 1.5 & 1.5  \\ % 20-30% of dmc/mc = 5%
%         \quad tuning $m_q$ & -- & -- & --  \\
%         \quad tuning $(m_l,m_h)$ & -- & -- & --  \\
%         \quad tuning $a$ & 0.5 & 2.4 & 2.5  \\ % ASK 6/26/05 from r1, 2S-1S
		higher-order $\rho_{A_4^{cq}}$ & 0 & 1.3 & 1.3 \\ % ASK 06/25/05
		heavy-quark discretization & 0.5 & 4.2 & 4.2 \\ % ASK 06/25/05
		light-quark discretization and $\chi$PT fits & 5.0 & 3.9 & 6.3 \\ % ASK 06/26/05
		static $\chi$PT & 1.4 & 0.5 & 1.5 \\ % CB 06/15/05, ASK 6/26/05
		finite volume & 1.4 & 0.5 & 1.5 \\ % CB 06/15/05, ASK 6/26/05
		\hline
		total systematic & 5.4 & 6.5 & 8.5 \\
		\hline\hline
	\end{tabular*}
\end{table}
all uncertainties are reducible in future work.
The results for $f_{D^+}$ and $f_{D_s}$ in Eqs.~(\ref{eq:fD+}) and
(\ref{eq:fDs}) are obtained via $f_{D_s}=\phi_s/\sqrt{m_{D_s}}$,
$f_{D^+}=R_{d/s}\phi_s/\sqrt{m_{D^+}}$,  by inserting the physical meson 
masses.

Present experimental measurements,
$f_{D^+} = 202 \pm 41 \pm 17~\textrm{MeV}$~\cite{Bonvicini:2004gv},
$f_{D_s} = 267 \pm 33~\textrm{MeV}$~\cite{Eidelman:wy},
are not yet precise enough to put our results in Eqs.~(\ref{eq:fD+})
and~(\ref{eq:fDs}) to a stringent test.  
The anticipated measurements of $f_{D^+}$ and, later,  $f_{D_s}$
from CLEO-c are therefore of great interest.
If validated, our calculation of $f_{D^+}$ has important implications 
for flavor physics.
For $B$ physics it is crucial to compute the decay constant~$f_B$.
To do so, we must simply change the heavy quark mass.
In fact, heavy-quark discretization effects, with the Fermilab method,
are expected to be smaller, about half as big.

We thank the U.S. National Science Foundation, the Office of Science of
the U.S. Department of Energy, Fermilab, and Indiana University for support, 
particularly for the computing needed for the project.
Fermilab is operated by Universities Research Association Inc., under
contract with the U.S. Department of Energy.

\emph{Note added:}~After this Letter was submitted, the CLEO-c
Collaboration announced a new measurement, 
$f_{D^+}=223\pm16^{+7}_{-9}~\textrm{MeV}$~\cite{Artuso:2005lp}.

\end{document}